\documentclass{aa}
\usepackage{psfig} 

\newcommand {\asec} {$^{\prime\prime}$}
\def\mic{$\,\mu\rm m$~}

\def\amin{\ifmmode ^{\prime}\else$^{\prime}$\fi}
\def\asec{\ifmmode ^{\prime\prime}\else$^{\prime\prime}$\fi}

\begin{document}
\thesaurus{03         
             (11.05.2;   
              13.09.1)   
}
\title{Source Counts from the 15\mic ISOCAM Deep Surveys
\thanks{Based on observations with ISO, an ESA project with instruments
funded by ESA Member States (especially the PI countries: France,
Germany, the Netherlands and the United Kingdom) with the participation
of ISAS and NASA} }  
\author{\bf D.~Elbaz\inst{1} 
\and C.J.~Cesarsky\inst{1,2}
\and D.~Fadda\inst{1}
\and H.~Aussel\inst{1,3}
\and F.X.~D\'esert\inst{4}
\and A.~Franceschini\inst{5}
\and H.~Flores\inst{1} 
\and M.~Harwit\inst{6}
\and J.L.~Puget\inst{7} 
\and J.L.~Starck\inst{1}
\and L.~Danese\inst{8} 
\and D.C.~Koo\inst{9}
\and R.~Mandolesi\inst{10}}
%
\offprints{delbaz@cea.fr}
\institute{
CEA Saclay - Service d'Astrophysique, Orme des Merisiers, 91191 Gif-sur-Yvette C\'edex, France \and
European Southern Observatory, Karl-Schwarzchild-Strasse, 2 D-85748 Garching bei Muenchen, Germany \and
Osservatorio di Padova, Vicolo Osservatorio, 5, I35122 Padova, Italy \and
Observatoire de Grenoble, BP 53, 414 rue de la piscine, 38041 Grenoble Cedex 9, France \and
Dipartimento di Astronomia, Universit\`a di Padova, Vicolo dell'Osservatorio, 5, I35122  Padova, Italy \and
511 H Street S.W., Washington, DC 20024-2725; also Cornell University, USA \and
Institut d'Astrophysique Spatiale, B{\^a}t 121, Universit\'e Paris XI, F-91405 Orsay C\'edex, France \and
SISSA, via Beirut 2-4, I-34013 Trieste, Italy \and
UCO/Lick Obs., University of California, Santa Cruz, CA 95064, USA \and
Bologna Radio Astronomy Institute, via Gobetti 101, 40129 Bologna, Italy.}

\date{Received: September 2, 1999; Accepted: October 17, 1999}

\maketitle
\authorrunning{D. Elbaz et al}
\titlerunning{Source Counts from ISOCAM Deep Surveys}
\markboth{D. Elbaz et al}{Source counts from ISOCAM deep surveys}

\begin{abstract}
We present the results of the five mid-IR 15\,$\mu$m (12-18\,$\mu$m
LW3 band) ISOCAM Guaranteed Time Extragalactic Surveys performed in
the regions of the Lockman Hole and Marano Field. The roughly 1000
sources detected, 600 of which have a flux above the 80~\%
completeness limit, guarantee a very high statistical significance for
the integral and differential source counts from 0.1\,mJy up to
$\sim$5\,mJy.  By adding the ISOCAM surveys of the HDF-North and South
(plus flanking fields) and the lensing cluster A2390 at low fluxes and
IRAS at high fluxes, we cover four decades in flux from 50\,$\mu$Jy to
$\sim$0.3\,Jy. The slope of the differential counts is very steep
($\alpha =-3.0$) in the flux range 0.4-4 mJy, hence much above the
Euclidean expectation of $\alpha =-2.5$. When compared with
no-evolution models based on IRAS, our counts show a factor $\sim$ 10
excess at 400\,$\mu$Jy, and a fast convergence, with $\alpha =-1.6$ at
lower fluxes.
\keywords{Galaxies: evolution -- 
Infrared: galaxies}
\end{abstract}
\section{Introduction}
Deep galaxy counts as a function of magnitude, or flux density,
should, in principle, give a constraint on the geometry of the
universe ($\Omega _{\circ}$, $\Lambda _{\circ}$). In fact, their
departure from the Euclidean expectation (no expansion, no curvature)
is dominated by the {\it e}-correction (intrinsic evolution of the
galaxies) and by the {\it k}-correction (redshift dependence). The
understanding of galaxy evolution therefore is a key problem for
cosmology, and number counts appear to be a strong constraint on the
models, which does not suffer from the exotic behavior of individual
galaxies. Most of the energy released by local galaxies is radiated in
the optical-UV range (Soifer \& Neugebauer 1991). If this were to
remain true over the whole history of the universe, then one could
follow the comoving star formation rate of the universe as a function
of redshift by measuring the optical-UV light radiated by galaxies
(Lilly et al 1996, Madau et al 1996). This scenario changed
considerably after the detection of a substantial diffuse cosmic IR
background (CIRB) in the 0.1 -- 1 mm range from the COBE-FIRAS data
(Puget et al 1996, Guiderdoni et al 1997, Hauser et al 1998, Fixsen
et al 1998, Lagache et al 1999) and at 140 -- 240 $\mu$m from the
COBE-DIRBE data (Hauser et al 1998, Lagache et al 1999). Surprisingly
the mid-IR/sub-mm extragalactic background light is at least as large
as that of the UV/optical/near-IR background (Dwek et al 1998, Lagache
et al 1999), which implies a stronger contribution of obscured star
formation at redshifts larger than those sampled by IRAS ($z>0.2$). To
understand the exact origin of this diffuse emission and its
implications for galaxy evolution, we need to identify the individual
galaxies responsible for it and the best way to do that consists of
observing directly in the IR/sub-mm range.

In the mid-IR, IRAS has explored the local universe ($z<0.2$).  With a
thousand times better sensitivity and sixty times better spatial
resolution, ISOCAM (Cesarsky et al 1996), the camera on-board ISO
(Kessler et al 1996), provides for the first time the opportunity to
perform cosmologically meaningful surveys.  Deep surveys have been
carried out on small fields containing sources well known at other
wavelengths: the HDF North (Serjeant et al 1997, Aussel et al 1999a,b,
D\'esert et al 1999) and the CFRS 1452+52 field (Flores et al
1999). This has yielded a small but meaningful sample of sources (83
galaxies) with a positional accuracy better than 6", sufficient for
most multiwavelength studies.  Most of these sources can easily be
identified with bright optical counterparts ($I_{C}<22.5$)
with a median redshift of $z\simeq0.7-0.8$ imposed by the {\it
k}-correction (Aussel et al 1999a,b, Flores et al 1999). Flores et al
(1999) estimate, from their sample of 41 sources, that accounting for
the IR light from star forming galaxies may lead to a global star
formation rate which is 2 to 3 times larger than estimated from UV
light only.

To obtain reliable source count diagrams, better statistics and a
wider range of flux densities are required. For this reason, we have
performed several cosmological surveys with ISOCAM, ranging from large
area-shallow surveys to small area-ultra deep surveys. These surveys
were obtained in the two main broad-band filters LW2 (5-8.5\,$\mu$m)
and LW3 (12-18\,$\mu$m), centered at 6.75\,$\mu$m and 15\,$\mu$m
respectively. This paper only presents the source counts at
15\,$\mu$m, because the sample of galaxies detected in the
6.75\,$\mu$m band is strongly contaminated by Galactic stars, whose
secure identification requires ground-based follow-up
observations. Including the surveys over the two Hubble deep fields,
almost 1000 sources with flux densities ranging from 0.1\,mJy to
10\,mJy were detected, allowing us to establish detailed source count
diagrams. Source lists for each individual survey, as well as maps and
detailed description of data reduction, will be found in separate
papers (see Table~\ref{TABLE:surveys}). Forthcoming papers will
discuss the nature of these galaxies based on the ISOHDF-North survey
(Aussel et al, in prep. \& 1999b), the contribution of these galaxies
to the cosmic IR background, the relation of these observations with
ISOPHOT and SCUBA deep surveys (Elbaz et al, in prep.), as well as a
tentative modelling of the number counts (Franceschini et al, in
prep.).
\section{Description of the Surveys} 
The five ISOCAM Guaranteed Time Extragalactic Surveys (Cesarsky \&
Elbaz 1996, Table~\ref{TABLE:surveys}) are complementary. They were
carried out in both the northern (Lockman Hole) and southern (Marano
Field) hemispheres, in order to be less biased by large-scale
structures. These two fields were selected for their low zodiacal and
cirrus emission and because they had been studied at other
wavelengths, in particular in the X-ray band, which is an indicator of
the AGN activity of the galaxies. Only one of the `Marano' maps was
scanned at the exact position of the original Marano Field (Marano,
Zamorani, Zitelli 1988), while the `Marano FIRBACK' (MFB) Deep and
Ultra-Deep surveys were positioned at the site of lowest galactic
cirrus emission, because they were combined with the FIRBACK ISOPHOT
survey at 175\,$\mu$m (Puget et al 1999, Dole et al 1999). Indeed the
importance of the Galactic cirrus emission in hampering source
detection is much larger at 175\,$\mu$m than at 15\,$\mu$m, but the
quality of the two 15\,$\mu$m ultra deep surveys in the Marano Field
area is equivalent.  In addition, very deep surveys were taken with
ISOCAM over the areas of the HDF North (Serjeant et al 1997) and HDF
South. In this paper, we include the HDF North results from Aussel
et al (1999a), and show for the first time ISOCAM counts on the HDF
South field.
\begin{table}[ht]
\caption{ISOCAM 15\,$\mu$m surveys sorted by increasing depth.}
\label{TABLE:surveys}
\footnotesize
\setlength{\tabcolsep}{0.6mm}
\begin{tabular}{lcccccc}
\hline
\multicolumn{1}{c}{Survey}&
{N$_{obs}$}&
{Area}&
{S$_{80\%}$}&
{$t_{int}$}&
{\# gal} &
{Slope}\\
\multicolumn{1}{c}{Name}&
&
{(am$^2$)}&
{(mJy)}&
{(min)}&
{} &
\\
\multicolumn{1}{c}{(1)}&
{(2)}&
{(3)}&
{(4)}&
{(5)}&
{(6)}&
{(7)}\\
\hline
Lockman Shallow$^{(a)}$ &  3&1944& 1   &   3&  80& $-2.1\pm0.2$\\
Lockman Deep$^{(a)}$&  6& 510& 0.6 &  11&  70& $-2.4\pm0.3$\\
MFB Deep$^{(a)}$    & 18& 710& 0.4 &15.4& 144& $-2.4\pm0.2$\\
Marano UD$^{(a)}$   & 75&  70& 0.2 & 114&  82& $-1.5\pm0.1$\\
MFB UD$^{(a)}$      & 75&  90& 0.2 & 114& 100& $-1.5\pm0.2$\\
HDF North$^{(b)}$   & 64&  27& 0.1 & 135&  44& $-1.6\pm0.2$\\
HDF South$^{(a)}$   & 64&  28& 0.1 & 168&  63& $-1.4\pm0.1$\\
A2390$^{(c)}$       &100& 5.3&0.05 & 432&  31& $-1.2\pm0.6$\\
\hline\\
\end{tabular}

\noindent{\footnotesize {\em Comments}: Col.(1) Survey name with reference: (a) in preparation, (b) Aussel et al (1999), (c) Altieri et al (1999); Col.(2) maximum number of pointings on the same sky position (redundancy); Col.(3) the total area covered in square arcminutes; Col.(4) the flux at which the survey is at least 80\% complete; Col.(5) the corresponding integration time per sky position (in minutes); Col.(6) the number of galaxies whose flux is over the 80\% completeness threshold; Col.(7) the slope of the fit to the integral  $\log N - \log S$. A2390 completeness limit includes the corrections for lensing magnification.}
\end{table}
\section{Data Reduction and Simulations}\label{data_reduction}
The transient behavior of the cosmic ray induced glitches, which makes
some of them mimic real sources, is the main limitation of ISOCAM
surveys.  We have developed two pipelines for the analysis of ISOCAM
surveys in order to obtain two independent source lists per survey and
improve the quality of the analysis. PRETI (Pattern REcognition
Technique for ISOCAM data), developed by Starck et al (1999), is able
to find and remove glitches using multi-resolution wavelet transforms.
\begin{figure}[h]
\psfig{figure=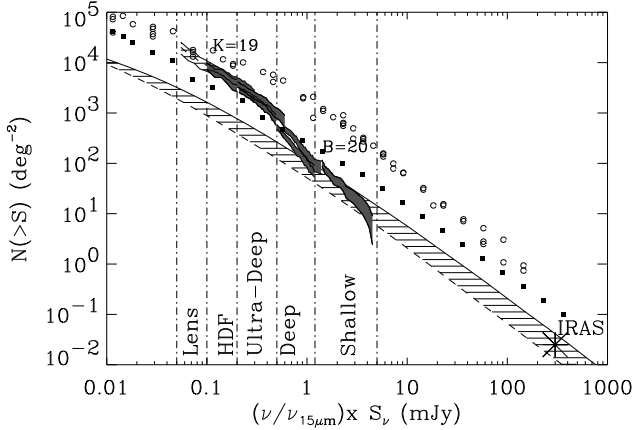,width=8.8cm}
\caption[]{Integral counts, i.e. the number of galaxies, N, detected
at 15\,$\mu$m above the flux S(mJy), with 68~\% confidence contours.
K counts (Gardner et al 1993) and B counts (Metcalfe et
al 1995),multiplied by the ratio $\nu/\nu_{15}$ to represent the
relative energy densities at high fluxes, are overplotted with open
circles and filled squares, respectively.  The hatched area
materializes the range of possible expectations from models assuming
no evolution and normalized to the IRAS 12\,$\mu$m local luminosity
function (LLF). The upper limit was calculated on the basis of the LLF
of Rush, Malkan \& Spinoglio (1993), as Xu et al (1998) and shifted
from 12 to 15\,$\mu$m with the template SED of M82; the lower limit
uses the LLF of Fang et al (1998) and the template SED of M51.}
\label{FIG:logN}
\end{figure}
It includes also Monte Carlo simulations to quantify the false
detection rate, to calibrate the photometry and to estimate the
completeness.  The `Triple Beam-Switch' (TBS) technique, developed by
D\'esert et al (1999), treats micro-scanning or mosaic images as if
they resulted from beam-switching observations.  All the surveys have
been independently analyzed using both techniques and the source lists
were cross-checked to attribute quality coefficients to the sources.
PRETI and TBS agree at the 20~$\%$ level in photometry (corresponding
to the photometric accuracy of both techniques), and with an
astrometric accuracy smaller than the pixel size (due to the
redundancy).  PRETI allowed us to attain fainter levels in deep
surveys, whereas in the shallow surveys, where the redundancy is not
very high, a very strict criterion of multiple detections had to be
applied. Finally, we have made Monte Carlo simulations by taking into
account the completeness and the photometric accuracy to correct for the
Eddington bias and to compute error bars in the number count plots.
\section{The ISOCAM 15 \mic source counts} 
\begin{figure}
\psfig{figure=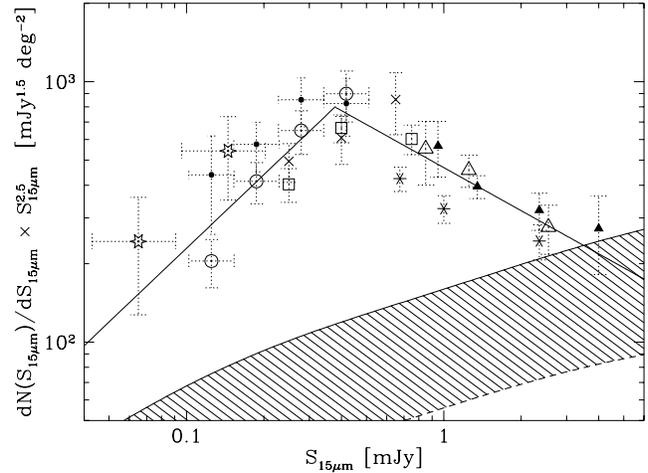,width=8.8cm}
\caption[]{Differential Number Counts of 15\,$\mu$m Galaxies, with
68\% error bars. The counts are normalized to a Euclidean distribution
of non-evolving sources, which would have a slope of $\alpha=-2.5$ in
such a universe.  Data points: A2390 (open stars), ISOHDF-North (open
circles), ISOHDF-South (filled circles), Marano FIRBACK (MFB)
Ultra-Deep (open squares), Marano Ultra-Deep (crosses), MFB Deep
(stars), Lockman Deep (open triangles), Lockman Shallow (filled
triangles).  The hatched area materializes the range of possible
expectations from models assuming no evolution (see Fig.1).}
\label{FIG:diff}
\end{figure}

Figures~\ref{FIG:logN} and~\ref{FIG:diff} show respectively the
integral and the differential number counts obtained in the five
independent guaranteed time surveys conducted in the ISOCAM 15\,$\mu$m
band,as well as the HDF surveys. The contribution of stars to the
15\,$\mu$m counts was corrected. It is negligible at fluxes below the
mJy level as confirmed by the spectro-photometric identifications in
the ISOHDF-North (1 star out of 44 sources), South (3 stars over 71
sources), and CFRS 1415+52 (1 star over 41 sources ranging from
$\sim 0.3$ mJy to $\sim 0.8$ mJy). In the Lockman Shallow Survey
($S_{15\,\mu m}\,> 1$ mJy), about 12~$\%$ of the sources were
classified as stars from their optical-mid IR colors (using the
Rayleigh-Jeans law). We have also represented the counts from the
ISOHDF-North (from Aussel et al 1999a), ISOHDF-South, and, at the
lowest fluxes, the counts obtained from the A2390 cluster lens (down
to 50\,$\mu$Jy, including the correction for lensing magnification;
Altieri et al 1999, see also Metcalfe et al 1999). We have only
included the flux bins where the surveys are at least 80~\% complete,
according to the simulations.

The first striking result of these complementary source counts is the
consistency of the eight 15\,$\mu$m surveys over the full flux
range. Some scatter is nevertheless apparent; given the small size of
the fields surveyed, we attribute it to clustering effects.
The two main features of the observed counts are a significantly
super-euclidean slope ($\alpha=-3.0$) from 3 to 0.4 mJy and a fast
convergence at flux densities fainter than 0.4 mJy. In particular, the
combination of five independent surveys in the flux range
90-400\,$\mu$Jy shows a turnover of the normalized differential counts
around 400\,$\mu$Jy and a decrease by a factor $\sim 3$ at
100\,$\mu$Jy.  We believe that this decrease, or the flattening of the
integral counts (see the change of slope in col(7) of
Table~\ref{TABLE:surveys}) below $\sim$400\,$\mu$Jy, is real.  It
cannot be due to incompleteness, since this has been estimated from
the Monte-Carlo simulations (see Section~\ref{data_reduction}).  The
differential counts can be fitted by two power laws by splitting the
flux axis in two regions around 0.4 mJy. In units of mJy$^{-1}$
deg$^{-2}$, we obtain, by taking into account the error bars ($S$ is in
mJy):
\begin{eqnarray}  
\frac{dN(S)}{dS} = \left\{ 
\begin{array}{cccc}
(2000\pm600)& S^{(-1.6\pm0.2)} &\ldots&  0.1\le S\le0.4\\
&&&\\
(470\pm30)& S^{(-3.0\pm0.1)} &\ldots& 0.4\le S\le4\\
\end{array}
\right.
\end{eqnarray} 
In the integral plot, the curves are plotted with 68~$\%$ confidence
contours based on our simulation analysis.  The total number density
of sources detected by ISOCAM at 15\,$\mu$m down to 100\,$\mu$Jy (no
lensing) is ($2.4\pm0.4$) arcmin$^{-2}$. It extends up to
($5.6\pm1.5$) arcmin$^{-2}$, down to 50\,$\mu$Jy, when including the
lensed field of A2390 (Altieri et al 1999).
%
\section{Discussion \& Conclusions} 
We have presented the 15\,$\mu$m differential and integral counts
drawn by several complementary ISOCAM deep surveys, with a significant
statistical sampling (993 galaxies, 614 of which have a flux above the
80~\% completeness limit) over two decades in flux from 50\,$\mu$Jy up
to 5\,mJy. 

The differential counts (Fig.~\ref{FIG:diff}), which are normalized to
$S^{-2.5}$ (the expected differential counts in a non expanding
Euclidean universe with sources that shine with constant luminosity),
present a turnover around $S_{15\,\mu m}$=0.4 mJy, above which the
slope is very steep ($\alpha=-3.0\pm0.1$).  No evolution predictions
were derived assuming a pure {\it k}-correction in a flat universe
($q_0=0.5$), including the effect of Unidentified Infrared Bands
emission in the galaxy spectra. In the
Fig.\ref{FIG:logN}\&\ref{FIG:diff}, the lower curve is based on the
Fang et al (1998) IRAS 12\,$\mu$m local luminosity function (LLF),
using the spectral template of a quiescent spiral galaxy (M51). The
upper curve is based on the Rush, Malkan \& Spinoglio (1993)
IRAS-12\,$\mu$m LLF, translated to 15\,$\mu$m using as template the
spectrum of M82, which is also typical of most starburst galaxies in
this band. More active and extincted galaxies, like Arp220, would lead
to even lower number counts at low fluxes while flatter spectra like
those of AGNs are less flat than M51. In the absence of a well
established LLF at 15\,$\mu$m, we consider these two models as upper
and lower bounds to the actual no-evolution expectations; note that
the corresponding slope is $\sim -2$. The actual number counts are
well above these predictions; in the 0.3\,mJy to 0.6\,mJy range, the
excess is around a factor 10: clearly, strong evolution is required to
explain this result (note the analogy with the radio source counts,
Windhorst et al 1993).
 
We believe, according to the results obtained on the HDF and CFRS
fields (Aussel et al 1999a,b, Flores et al 1999), that the sources
responsible for the 'bump' in the 15\,$\mu$m counts are not the faint
blue galaxies which dominate optical counts at $z\sim0.7$.  Instead,
they most probably are bright and massive galaxies whose emission
occurs essentially in the IR and could account for a considerable part
of the star formation in the universe at $z<1.5$.

In Fig.~\ref{FIG:logN}, we have overplotted the integral counts in the
K (Gardner et al 1993) and B (Metcalfe et al 1995) bands, in terms of
$\nu S_{\nu}$.  For bright sources, with densities lower than 10
deg$^{-2}$, these curves run parallel to an interpolation between the
ISOCAM counts presented here and the IRAS counts; the bright K sources
emit about ten times more energy in this band than a comparable number
of bright ISOCAM sources at 15 \mic. But the ISOCAM integral counts
present a rapid change of slope around 1-2 mJy, and their numbers rise
much faster than those of the K and B sources. The sources detected by
ISOCAM are a sub-class of the K and B sources which harbor activity
hidden by dust. Linking luminosity to distance, we predict a rapid
change of the luminosity function with increasing redshift, which can
only be confirmed by a complete ground-based spectro-photometric
follow-up. We should be able to follow the evolution of the luminosity
function from $z\sim0.2$ to $1.5$ with the large number of galaxies
detected in the Marano Field surveys.  The combination of the
intensity of the $H_{\alpha}$ emission line (redshifted in the J-band)
with the IR luminosity could set strong constraints on the star
formation rate. Finally, emission line diagnostics, combined with hard
X-ray observations with XMM and Chandra, would allow us to understand
whether the dominant source of energy is star formation or AGN
activity.

\begin{acknowledgements}
 One of us (MH) wishes to acknowledge the hospitality of the Max
Planck Inst. for Radioastronomy and the A. von Humboldt Foundation of
Germany during work on this paper; his research with ISO is funded by
a NASA grant.
\end{acknowledgements}

\end{document}